\documentclass{llncs}

\usepackage{amsfonts}
\usepackage{amsmath}
\usepackage{amssymb}
\usepackage{latexsym}
\usepackage{color}
\usepackage{graphics}
\usepackage{enumerate}
\usepackage{psfig}
\usepackage{amstext}
\usepackage{pseudocode}
\usepackage[whileod]{algochl}
\usepackage{gastex}
\usepackage{url}
\usepackage{ifthen}
\usepackage{times}

\def\Kset{\mathbb{K}}

\def\Rset{\mathbb{R}}

\newcommand{\bi}{\begin{itemize}}
\newcommand{\ei}{\end{itemize}}
\newcommand{\be}{\begin{enumerate}}
\newcommand{\ee}{\end{enumerate}}
\newcommand{\bd}{\begin{description}}
\newcommand{\ed}{\end{description}}

\newcommand{\T}{(\Rset_+ \cup \{\infty\}, \min, +, \infty, 0)}

\newcommand{\Trans}{T = (\Sigma, \Delta, Q, I, F, E, \lambda, \rho)}

\newcommand{\den}[1]{[\![ #1 ]\!]}
\newcommand{\set}[1]{\{#1\}}

\newcommand{\ipsfig}[2]{\scalebox{#1}{\psfig{#2}}}

\newcommand{\ignore}[1]{}

\newcommand{\e}{\epsilon}
\newcommand{\+}{\oplus}
\renewcommand{\.}{\otimes}
\newcommand{\0}{\overline{0}}

\newcommand{\nsp}{\!\!}
\newcommand{\coln}{\nsp:\nsp}

\title{3-Way Composition \\of Weighted Finite-State Transducers}  

\author{
Cyril Allauzen \inst{1} \fnmsep \thanks{This author's current address is: Google Research, 76 Ninth Avenue, New York, NY 10011.}
\and Mehryar Mohri\inst{1, 2}
\institute{
Courant Institute of Mathematical Sciences,\\
251 Mercer Street, New York, NY 10012.
\and
Google Research,\\
76 Ninth Avenue, New York, NY 10011.}}

\begin{document}

\maketitle

\begin{abstract}
Composition of weighted transducers is a fundamental algorithm used in
many applications, including for computing complex edit-distances
between automata, or string kernels in machine learning, or to combine
different components of a speech recognition, speech synthesis, or
information extraction system. We present a generalization of the
composition of weighted transducers, \emph{$3$-way composition}, which
is dramatically faster in practice than the standard composition
algorithm when combining more than two transducers. The worst-case
complexity of our algorithm for composing three transducers $T_1$,
$T_2$, and $T_3$ resulting in $T$, \ignore{ depending on the strategy
used, is $O(|T|_Q d(T_1) d(T_3) + |T|_E)$ or $(|T|_Q d(T_2) + |T|_E)$,
} is $O(|T|_Q \min(d(T_1) d(T_3), d(T_2)) + |T|_E)$, where $|\cdot|_Q$
denotes the number of states, $|\cdot|_E$ the number of transitions,
and $d(\cdot)$ the maximum out-degree. As in regular composition, the
use of perfect hashing requires a pre-processing step with linear-time
expected complexity in the size of the input transducers. In many
cases, this approach significantly improves on the complexity of
standard composition. Our algorithm also leads to a dramatically
faster composition in practice. Furthermore, standard composition can
be obtained as a special case of our algorithm. We report the results
of several experiments demonstrating this improvement. These
theoretical and empirical improvements significantly enhance
performance in the applications already mentioned.
\end{abstract}

\section{Introduction}

Weighted finite-state transducers are widely used in text, speech, and
image processing applications and other related areas such as
information extraction
\cite{cl,lothaire,pereira-riley,ecai,Culik1997}. They are finite
automata in which each transition is augmented with an output label
and some weight, in addition to the familiar (input) label
\cite{soittola,eilenberg,kuich}. The weights may represent
probabilities, log-likelihoods, or they may be some other costs used
to rank alternatives. They are, more generally, elements of a semiring
\cite{kuich}.

Weighted transducers are used to represent models derived from large
data sets using various statistical learning techniques such as
pronunciation dictionaries, statistical grammars, string kernels, or
complex edit-distance models \cite{ecai,Katz87,Chen98,jmlr}. These
models can be combined to create complex systems such as a speech
recognition or information extraction system using a fundamental
transducer algorithm, \emph{composition of weighted transducers}
\cite{pereira-riley,ecai}. Weighted composition is a generalization of
the composition algorithm for unweighted finite-state transducers
which consists of matching the output label of the transitions of one
transducer with the input label of the transitions of another
transducer. The weighted case is however more complex and requires the
introduction of an $\e$-filter to avoid the creation of redundant
$\e$-paths and preserve the correct path multiplicity
\cite{pereira-riley,ecai}. The result is a new weighted transducer
representing the relational composition of the two transducers.

Composition is widely used in computational biology, text and speech,
and machine learning applications. In many of these applications, the
transducers used are quite large, they may have as many as several
hundred million states or transitions. A critical problem is thus to
devise efficient algorithms for combining them.
This paper presents a generalization of the composition of weighted
transducer, \emph{$3$-way composition}, that is dramatically faster
than the standard composition algorithm when combining more than two
transducers. The complexity of composing three transducer
$T_1$, $T_2$, and $T_3$, with the standard composition algorithm is
$O(|T_1||T_2||T_3|)$ \cite{pereira-riley,ecai}. Using perfect hashing,
the worst-case complexity of computing $T = (T_1 \circ T_2)
\circ T_3$ using standard composition is
\begin{equation} 
\small O(|T|_Q \min(d(T_3), d(T_1 \circ T_2)) + |T|_E
+ |T_1 \circ T_2|_Q \min(d(T_1), d(T_2)) + |T_1 \circ T_2|_E),
\end{equation} 
which may be prohibitive in some cases even when the resulting
transducer $T$ is not large but the intermediate transducer $T_1 \circ
T_2$ is. Instead, the worst-case complexity of our algorithm is
\begin{equation}
\small O(|T|_Q \min(d(T_1) d(T_3), d(T_2)) + |T|_E).
\end{equation}
In both cases, the use of perfect hashing requires a pre-processing
step with linear-time expected complexity in the size of the input
transducers. 

Our algorithm also leads to a dramatically faster computation of the
result of composition in practice. We report the results of several
experiments demonstrating this improvement. These theoretical and
empirical improvements significantly enhance performance in a series
of applications: string kernel-based algorithms in machine learning,
the computation of complex edit-distances between automata, speech
recognition and speech synthesis, and information
extraction. Furthermore, as we shall see later, standard composition
can be obtained as a special case of $3$-way composition.

The main technical difficulty in the design of our algorithm is the
definition of a \emph{filter} to deal with a path multiplicity problem
that arises in the presence of the empty string $\e$ in the
composition of three transducers. This problem, which we shall
describe in detail, leads to a word combinatorial problem
\cite{perrin-lothaire}. We will present two solutions for this
problem: one requiring two $\e$-filters and a generalization of the
$\e$-filters used for standard composition \cite{pereira-riley,ecai};
and another direct and symmetric solution where a single filter is
needed. Remarkably, this $3$-way filter can be encoded as a finite
automaton and painlessly integrated in our $3$-way composition.

The remainder of the paper is structured as follows. Some preliminary
definitions and terminology are introduced in the next section
(Section~\ref{sec:preliminaries}). Section~\ref{sec:algorithm}
describes our $3$-way algorithm in the $\e$-free case. The word
combinatorial problem of $\e$-path multiplicity and our solutions are
presented in detail Section~\ref{sec:epsilon filtering}. 
Section~\ref{sec:experiments} reports the results of experiments using
the $3$-way algorithm and compares them with the standard composition.

\section{Preliminaries}
\label{sec:preliminaries}

This section gives the standard definition and specifies the notation
used for weighted transducers. 

\emph{Finite-state transducers} are finite automata in which each
transition is augmented with an output label in addition to the
familiar input label \cite{berstel,eilenberg}. Output labels are
concatenated along a path to form an output sequence and similarly
with input labels. \emph{Weighted transducers} are finite-state
transducers in which each transition carries some weight in addition
to the input and output labels \cite{soittola,kuich}. 

The weights are elements of a semiring, that is a ring that may lack
negation \cite{kuich}. Some familiar semirings are the tropical
semiring $\T$ related to classical shortest-paths algorithms, and the
probability semiring $(\Rset, +, \cdot , 0, 1)$. A semiring is
\emph{idempotent} if for all $a \in \Kset$, $a \oplus a = a$. It is
\emph{commutative} when $\.$ is commutative. We will assume in this
paper that the semiring used is commutative, which is a necessary
condition for composition to be an efficient algorithm
\cite{lothaire}.

The following gives a formal definition of weighted transducers.

\begin{definition}
A {\em weighted finite-state transducer} $T$ over $(\Kset, \+, \cdot,
0, 1)$ is an 8-tuple $\Trans$ where $\Sigma$ is the finite input
alphabet of the transducer, $\Delta$ is the finite output alphabet,
$Q$ is a finite set of states, $I \subseteq Q$ the set of initial
states, $F \subseteq Q$ the set of final states, $E \subseteq Q \times
(\Sigma \cup \set{\epsilon}) \times (\Delta \cup \set{\epsilon})
\times \Kset \times Q$ a finite set of transitions, $\lambda: I
\rightarrow \Kset$ the initial weight function, and $\rho: F
\rightarrow \Kset$ the final weight function mapping $F$ to $\Kset$.

\end{definition}
The weight of a path $\pi$ is obtained by multiplying the weights of
its constituent transitions using the multiplication rule of the
semiring and is denoted by $w[\pi]$. The weight of a pair of input and
output strings $(x, y)$ is obtained by $\+$-summing the weights of the
paths labeled with $(x, y)$ from an initial state to a final state.

\begin{figure*}[t]
\begin{center}
\begin{tabular}{c@{\hspace{2cm}}c}
\ipsfig{.4}{figure=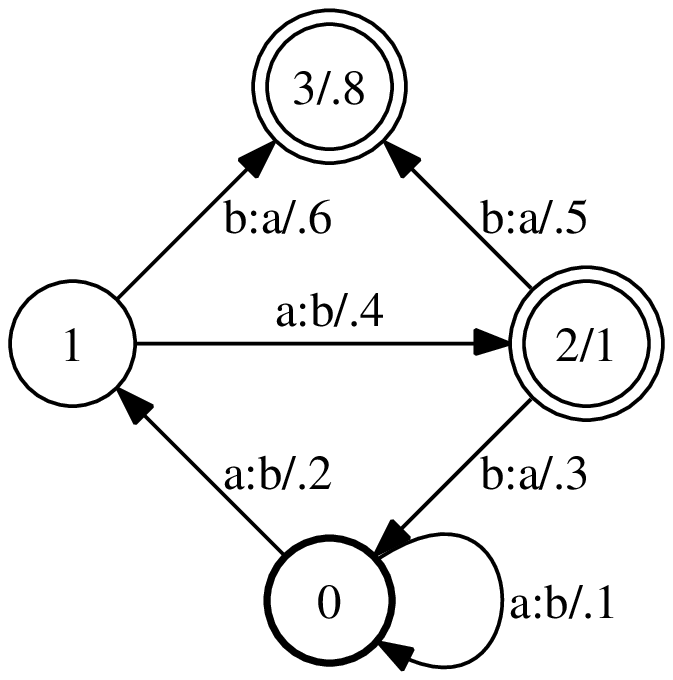} & \ipsfig{.4}{figure=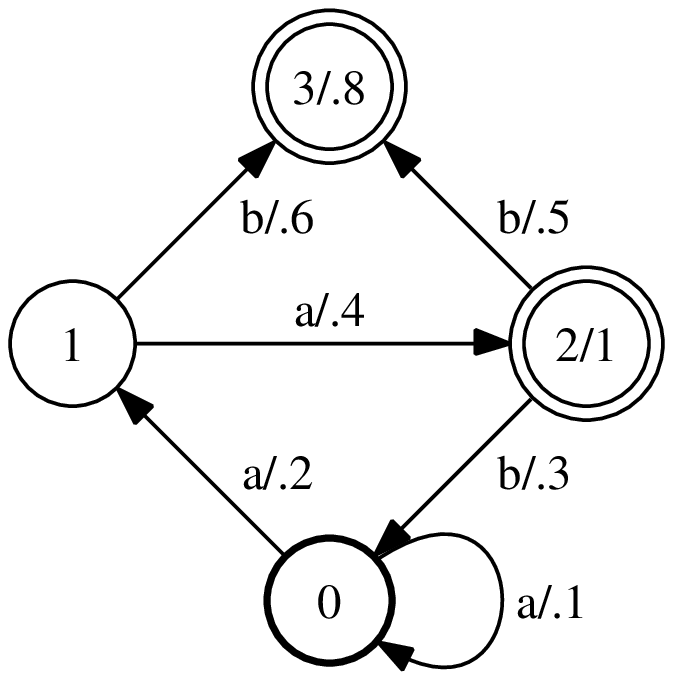}\\
(a) & (b)
\end{tabular}
\end{center}
\caption[]{(a) Example of a weighted transducer $T$. (b) Example of a
weighted automaton $A$. $\den{T}(aab, bba) = \den{A}(aab) = .1 \times
.2 \times .6 \times .8 + .2 \times .4 \times .5 \times .8$. A bold
circle indicates an initial state and a double-circle a final state.
The final weight $\rho[q]$ of a final state $q$ is indicated after the
slash symbol representing $q$. }
\label{fig:examples}
\end{figure*}

For a path $\pi$, we denote by $p[\pi]$ its origin state and by
$n[\pi]$ its destination state. We also denote by $P(I, x, y, F)$ the
set of paths from the initial states $I$ to the final states $F$
labeled with input string $x$ and output string $y$. A transducer $T$
is {\em regulated} if the output weight associated by $T$ to any pair
of strings $(x, y)$:
\begin{equation}
T(x, y) = \bigoplus_{\pi \in P(I, x, y, F)} \lambda(p[\pi])
\cdot w[\pi] \cdot \rho[n[\pi]]
\end{equation}
is well-defined and in $\Kset$. $T(x, y) = \0$ when $P(I, x, y, F) =
\emptyset$. If for all $q \in Q$ $\bigoplus_{\pi \in P(q, \epsilon,
\epsilon, q)} w[\pi] \in \Kset$, then $T$ is regulated. In particular,
when $T$ does not admit any $\epsilon$-cycle, it is regulated. The
weighted transducers we will be considering in this paper will be
regulated. Figure~\ref{fig:examples}(a) shows an example.

The {\em composition} of two weighted transducers $T_1$ and $T_2$ with
matching input and output alphabets $\Sigma$, is a weighted transducer
denoted by $T_1 \circ T_2$ when the sum:
\begin{equation}
(T_1 \circ T_2)(x, y) = \bigoplus_{z \in \Sigma^*}\ T_1(x,
z) \. T_2(z, y)
\end{equation}
is well-defined and in $\Kset$ for all $x, y \in \Sigma^*$
\cite{soittola,kuich}. \emph{Weighted automata} can be defined as
weighted transducers $A$ with identical input and output labels, for
any transition. Thus, only pairs of the form $(x, x)$ can have a
non-zero weight by $A$, which is why the weight associated by $A$ to
$(x, x)$ is abusively denoted by $A(x)$ and identified with the
\emph{weight associated by $A$ to $x$}.  Similarly, in the graph
representation of weighted automata, the output (or input) label is
omitted.

\section{Epsilon-Free Composition}
\label{sec:algorithm}

\subsection{Standard Composition}

\begin{figure}[t]
\begin{center}
\begin{tabular*}{\textwidth}{@{\hspace{0cm}}c@{\extracolsep{\fill}}c@{\extracolsep{\fill}}c@{\hspace{0cm}}}
\ipsfig{.23}{figure=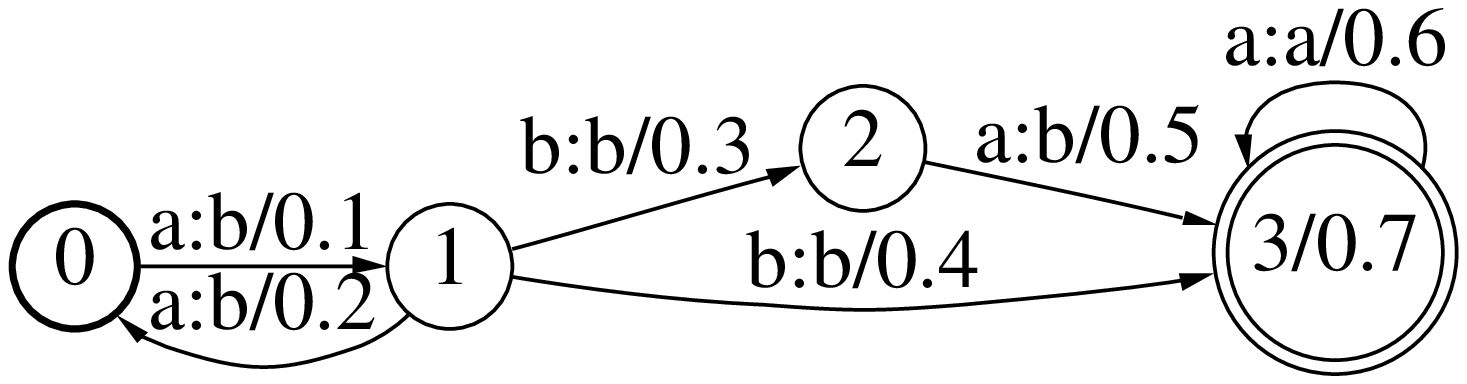} & \ipsfig{.23}{figure=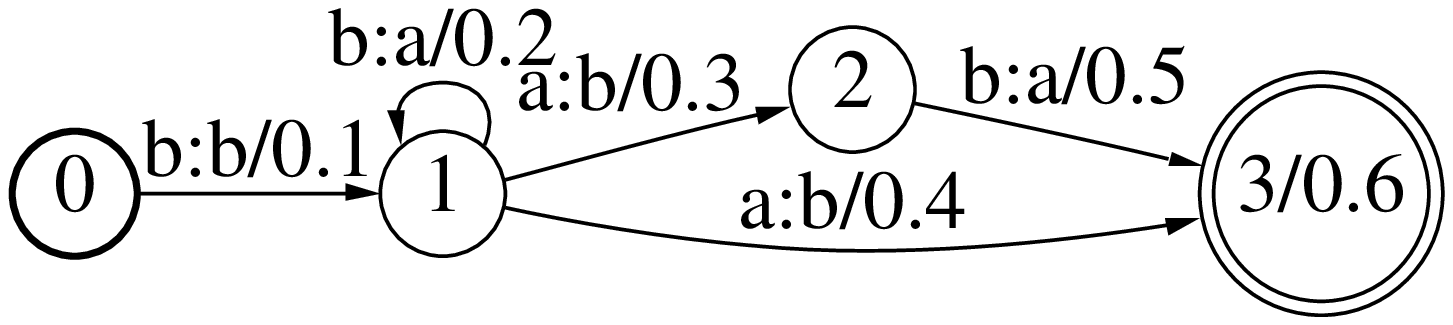} & \raisebox{-.4cm}{\ipsfig{.23}{figure=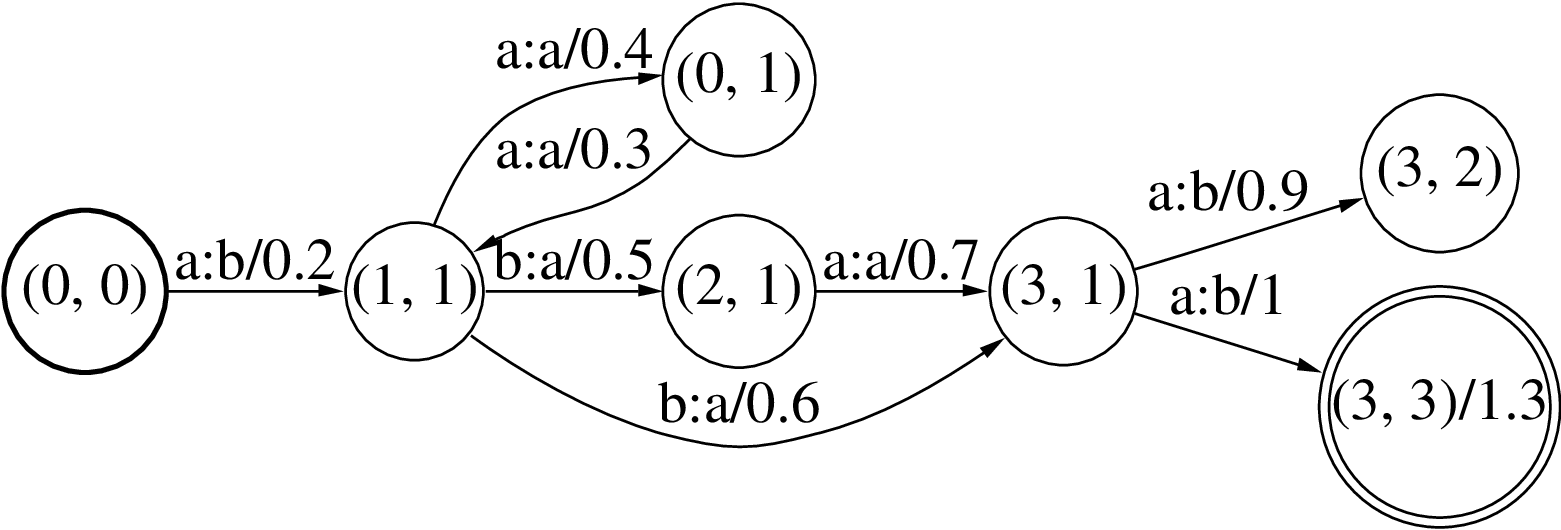}}\\
{\small(a)} & {\small (b)} & {\small(c)}\\
\end{tabular*}
\end{center}
\caption{Example of transducer composition.  (a) Weighted transducer
$T_1$ and (b) Weighted transducer $T_2$ over the probability semiring
$(\Rset, +, \cdot , 0, 1)$. (c) Result of the composition of $T_1$ and
$T_2$.}
\label{fig:composition}
\end{figure}

Let us start with a brief description of the standard composition
algorithm for weighted transducers \cite{pereira-riley,ecai}.  States
in the composition $T_1 \circ T_2$ of two weighted transducers $T_1$
and $T_2$ are identified with pairs of a state of $T_1$ and a state of
$T_2$. Leaving aside transitions with $\e$ inputs or outputs, the
following rule specifies how to compute a transition of $T_1 \circ
T_2$ from appropriate transitions of $T_1$ and $T_2$:
\begin{equation}
(q_1, a, b, w_1, q_2) \mbox{ and } (q'_1, b, c, w_2, q'_2)
\Longrightarrow ((q_1,q'_1), a, c, w_1 \. w_2, (q_2,q'_2)).
\end{equation}
Figure~\ref{fig:composition} illustrates the algorithm. In the worst
case, all transitions of $T_1$ leaving a state $q_1$ match all those
of $T_2$ leaving state $q'_1$, thus the space and time complexity of
composition is quadratic: $O(|T_1||T_2|)$. However, using perfect
hashing on the input transducer with the highest out-degree leads to a
worst-case complexity of
$O(|T_1 \circ T_2|_Q \min(d(T_1), d(T_2)) + |T_1 \circ T_2|_E)$.
The pre-processing step required for hashing the transitions
of the transducer with the highest out-degree has an expected
complexity in $O(|T_1|_E)$ if $d(T_1) > d(T_2)$ and $O(|T_2|_E)$
otherwise.

The main problem with the standard composition algorithm is the
following.  Assume that one wishes to compute $T_1 \circ T_2 \circ
T_3$, say for example by proceeding left to right. Thus, first $T_1$
and $T_2$ are composed to compute $T_1 \circ T_2$ and then the result
is composed with $T_3$. The worst-case complexity of that
computation is:
\begin{align}
O(& |T_1 \circ T_2 \circ T_3|_Q \min(d(T_1 \circ T_2), d(T_3)) 
   + |T_1 \circ T_2 \circ T_3|_E + \nonumber\\ 
  & |T_1 \circ T_2|_Q \min(d(T_1), d(T_2)) + |T_1 \circ T_2|_E).
\end{align}
But, in many cases, computing $T_1 \circ T_2$ creates a very large
number of transitions that may never match any transition of $T_3$.
For example, $T_2$ may represent a complex edit-distance transducer,
allowing all possible insertions, deletions, substitutions and perhaps
other operations such as transpositions or more complex edits in $T_1$
all with different costs. Even when $T_1$ is a simple
non-deterministic finite automaton with $\e$-transitions, which is
often the case in the applications already mentioned, $T_1 \circ T_2$
will then have a very large number of paths, most of which will not
match those of the non-deterministic automaton $T_3$. In other
applications in speech recognition, or for the computation of kernels
in machine learning, the central transducer $T_2$ could be far more
complex and the set of transitions or paths of $T_1 \circ T_2$ not
matching those of $T_3$ could be even larger.

\subsection{3-Way Composition}

The key idea behind our algorithm is precisely to avoid creating these
unnecessary transitions by directly constructing $T_1 \circ T_2 \circ
T_3$, which we refer to as a \emph{$3$-way composition}. Thus, our
algorithm does not include the intermediate step of creating $T_1
\circ T_2$ or $T_2 \circ T_3$. To do so, we can proceed following a
\emph{lateral} or \emph{sideways strategy}: for each transition $e_1$
in $T_1$ and $e_3$ in $T_3$, we search for matching transitions in
$T_2$. 

The pseudocode of the algorithm in the $\e$-free case is given below.
The algorithm computes $T$, the result of the composition $T_1 \circ
T_2 \circ T_3$. It uses a queue $S$ containing the set of pairs of
states yet to be examined. The queue discipline of $S$ can be
arbitrarily chosen and does not affect the termination of the
algorithm. Using a FIFO or LIFO discipline, the queue operations can
be performed in constant time.  We can pre-process the transducer
$T_2$ in expected linear time $O(|T_2|_E)$ by using perfect hashing so
that the transitions $G$ (line 13) can be found in worst-case linear
time $O(|G|)$. Thus, the worst-case running time complexity of the
$3$-way composition algorithm is in $O(|T|_Q d(T_1) d(T_3) + |T|_E)$,
where $T$ is transducer returned by the algorithm.

Alternatively, depending on the size of the three transducers, it may
be advantageous to direct the $3$-way composition from the center,
i.e., ask for each transition $e_2$ in $T_2$ if there are matching
transitions $e_1$ in $T_1$ and $e_3$ in $T_3$. We refer to this as the
\emph{central strategy} for our $3$-way composition
algorithm. Pre-processing the transducers $T_1$ and $T_3$ and creating
hash tables for the transitions leaving each state (the expected
complexity of this pre-processing being $O(|T_1|_E + |T_3|_E)$), this
strategy leads to a worst-case running time complexity of $O(|T|_Q
d(T_2) + |T|_E)$. The lateral and central strategies can be combined
by using, at a state $(q_1, q_2, q_3)$, the lateral strategy if
$|E[q_1]| \cdot |E[q_3]| \le |E[q_2]$ and the central strategy
otherwise. The algorithm leads to a natural lazy or on-demand
implementation in which the transitions of the resulting transducer
$T$ are generated only as needed by other operations on $T$. The
standard composition coincides with the $3$-way algorithm when using
the central strategy with either $T_1$ or $T_2$ equal to the identity
transducer.

{\small
\begin{ALGO}{3-Way-Composition}{T_1, T_2, T_3}
\SET{Q}{I_1 \times I_2 \times I_3}
\SET{S}{I_1 \times I_2 \times I_3}
\DOWHILE{S \neq \emptyset}
  \SET{(q_1, q_2, q_3)}{\Call{Head}{S}}
  \CALL{Dequeue}{S}
  \IF{(q_1, q_2 ,q_3) \in I_1 \times I_2 \times I_3} 
     \SET{I}{I \cup \set{(q_1, q_2, q_3)}}
     \SET{\lambda{(q_1, q_2, q_3)}}{\lambda_1(q_1) \. \lambda_2(q_2) \. \lambda_3(q_3)}
  \FI
  \IF{(q_1, q_2, q_3) \in F_1 \times F_2 \times F_3} 
     \SET{F}{F \cup \set{(q_1, q_2, q_3)}}
     \SET{\rho{(q_1, q_2, q_3)}}{\rho_1(q_1) \. \rho_2(q_2) \. \rho_3(q_3)}
  \FI
  \DOFOR{\tm{each } (e_1, e_3) \in E[q_1] \times E[q_3]}
    \SET{G}{\set{e \in E[q_2]: i[e] = o[e_1] \wedge o[e] = i[e_3]}}
    \DOFOR{\tm{each } e_2 \in G}
    \IF{(n[e_1], n[e_2], n[e_3]) \not \in Q}
      \SET{Q}{Q \cup \set{(n[e_1], n[e_2], n[e_3])}}
      \CALL{Enqueue}{S, (n[e_1], n[e_2], n[e_3])}
    \FI
    \OD
  \SET{E}{E \cup \set{((q_1, q_2, q_3), i[e_1], o[e_3], w[e_1] \.
  w[e_2] \. w[e_3], (n[e_1], n[e_2], n[e_3]))}}
  \OD
\OD
\RETURN{T}
\end{ALGO}}

\section{Epsilon filtering}
\label{sec:epsilon filtering}

The algorithm described thus far cannot be readily used in most cases
found in practice. In general, a transducer $T_1$ may have transitions
with output label $\e$ and $T_2$ transitions with input $\e$. A
straightforward generalization of the $\e$-free case would generate
redundant $\e$-paths and, in the case of non-idempotent semirings,
would lead to an incorrect result, even just for composing two
transducers. The weight of two matching $\e$-paths of the original
transducers would be counted as many times as the number of redundant
$\e$-paths generated in the result, instead of one. Thus, a crucial
component of our algorithm consists of coping with this problem.

\begin{figure}[t]
\begin{center}
\begin{tabular}{cc}
\scalebox{.75}{\newcommand{\radius}{8}
\newcommand{\xa}{0}
\newcommand{\xb}{20}
\newcommand{\xc}{40}

\newcommand{\ya}{40}
\newcommand{\yb}{20}
\newcommand{\yc}{0}

\begin{picture}(60,50)(0,0)

\node[Nw=\radius,Nh=\radius,Nmr=\radius](00)(\xa,\ya){$(0,\!0)$}
\node[Nw=\radius,Nh=\radius,Nmr=\radius](01)(\xa,\yb){$(0,\!1)$}
\node[Nw=\radius,Nh=\radius,Nmr=\radius](02)(\xa,\yc){$(0,\!2)$}

\node[Nw=\radius,Nh=\radius,Nmr=\radius](10)(\xb,\ya){$(1,\!0)$}
\node[Nw=\radius,Nh=\radius,Nmr=\radius](11)(\xb,\yb){$(1,\!1)$}
\node[Nw=\radius,Nh=\radius,Nmr=\radius](12)(\xb,\yc){$(1,\!2)$}

\node[Nw=\radius,Nh=\radius,Nmr=\radius](20)(\xc,\ya){$(2,\!0)$}
\node[Nw=\radius,Nh=\radius,Nmr=\radius](21)(\xc,\yb){$(2,\!1)$}
\node[Nw=\radius,Nh=\radius,Nmr=\radius](22)(\xc,\yc){$(2,\!2)$}

\drawedge[curvedepth=0](00,10){$\e_1{:}\e_1$}
\drawedge[curvedepth=0](01,11){$\e_1{:}\e_1$}
\drawedge[curvedepth=0](02,12){$\e_1{:}\e_1$}

\drawedge[curvedepth=0](10,20){$\e_1{:}\e_1$}
\drawedge[curvedepth=0](11,21){$\e_1{:}\e_1$}
\drawedge[curvedepth=0](12,22){$\e_1{:}\e_1$}

\drawedge[curvedepth=0,ELside=r](00,01){$\e_2{:}\e_2$}
\drawedge[curvedepth=0,ELside=r](01,02){$\e_2{:}\e_2$}

\drawedge[curvedepth=0](10,11){$\e_2{:}\e_2$}
\drawedge[curvedepth=0,linewidth=.4](11,12){$\e_2{:}\e_2$}

\drawedge[curvedepth=0,ELside=l](20,21){$\e_2{:}\e_2$}
\drawedge[curvedepth=0,ELside=l](21,22){$\e_2{:}\e_2$}

\drawedge[curvedepth=0,ELpos=52,linewidth=.4,ELdist=0](00,11){$\e_2{:}\e_1$}
\drawedge[curvedepth=0,ELpos=52,ELdist=0](01,12){$\e_2{:}\e_1$}
\drawedge[curvedepth=0,ELpos=52,ELdist=0](10,21){$\e_2{:}\e_1$}
\drawedge[curvedepth=0,ELpos=52,ELdist=0](11,22){$\e_2{:}\e_1$}

\end{picture} } & \ipsfig{.5}{figure=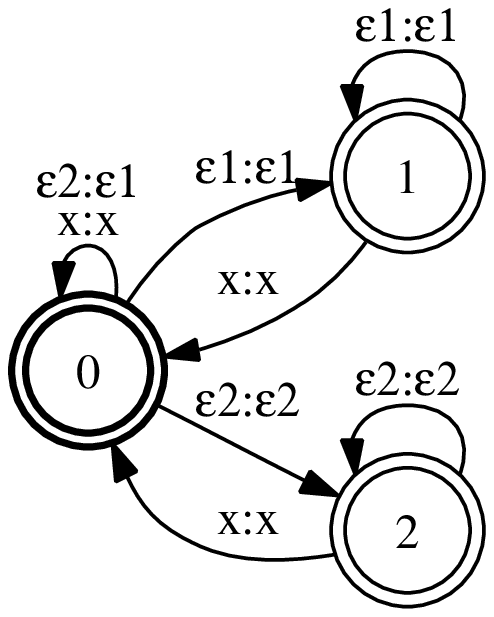}\\
(a) & (b)
\end{tabular}
\end{center}
\caption{(a) Redundant $\e$-paths. A straightforward generalization of
the $\e$-free case could generate all the paths from $(0, 0)$ to $(2,
2)$ for example, even when composing just two simple transducers. (b)
Filter transducer $M$ allowing a unique $\e$-path.}
\label{fig:e-paths}
\end{figure}

Figure~\ref{fig:e-paths}(a) illustrates the problem just mentioned in
the simpler case of two transducers. To match $\e$-paths leaving $q_1$
and those leaving $q_2$, a generalization of the $\e$-free composition
can make the following moves: (1) first move forward on a transition
of $q_1$ with output $\e$, or even a path with output $\e$, and stay
at the same state $q_2$ in $T_2$, with the hope of later finding a
transition whose output label is some label $a \neq \e$ matching a
transition of $q_2$ with the same input label; (2) proceed similarly
by following a transition or path leaving $q_2$ with input label $\e$
while staying at the same state $q_1$ in $T_1$; or, (3) match a
transition of $q_1$ with output label $\e$ with a transition of $q_2$
with input label $\e$.

Let us rename existing output $\e$-labels of $T_1$ as $\e_2$, and
existing input $\e$-labels of $T_2$ $\e_1$, and let us augment $T_1$
with a self-loop labeled with $\e_1$ at all states and similarly,
augment $T_2$ with a self-loop labeled with $\e_2$ at all states, as
illustrated by Figures~\ref{fig:marking}(a) and (c). These self-loops
correspond to staying at the same state in that machine while
consuming an $\e$-label of the other transition. The three moves just
described now correspond to the matches (1) $(\e_2 \coln \e_2)$,
(2) $(\e_1 \coln \e_1)$, and (3) $(\e_2 \coln \e_1)$. The grid
of Figure~\ref{fig:e-paths}(a) shows all the possible $\e$-paths
between composition states.  We will denote by $\tilde{T}_1$ and
$\tilde{T}_2$ the transducers obtained after application of these
changes.

For the result of composition to be correct, between any two of these
states, all but one path must be disallowed. There are many possible
ways of selecting that path. One natural way is to select the
shortest path with the diagonal transitions ($\e$-matching
transitions) taken first. Figure~\ref{fig:e-paths}(a) illustrates in
boldface the path just described from state $(0, 0)$ to state $(1,
2)$. Remarkably, this filtering mechanism itself can be encoded as a
finite-state transducer such as the transducer $M$ of
Figure~\ref{fig:e-paths}(b). We denote by $(p, q) \preceq (r, s)$ to
indicate that $(r, s)$ can be reached from $(p, q)$ in the grid.

\begin{proposition}
\label{prop:prop1}
Let $M$ be the transducer of Figure~\ref{fig:e-paths}(b). $M$ allows a
unique path between any two states $(p, q)$ and $(r, s)$, with $(p, q)
\preceq (r, s)$.
\end{proposition}

\begin{figure}[t]
\begin{center}
\begin{tabular*}{.8\textwidth}{@{\extracolsep{\fill}}ccc}
\ipsfig{.4}{figure=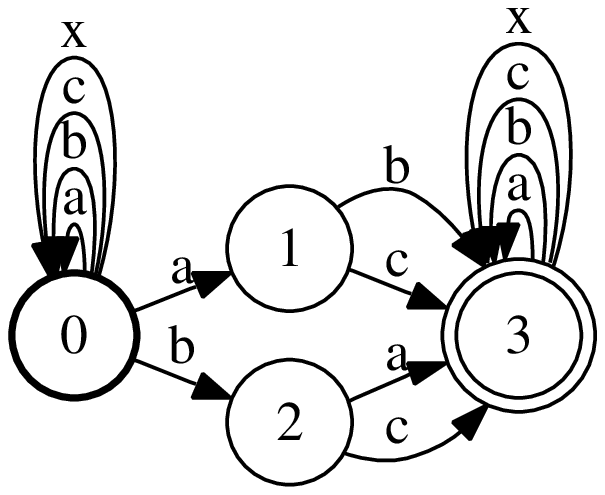}&
\ipsfig{.4}{figure=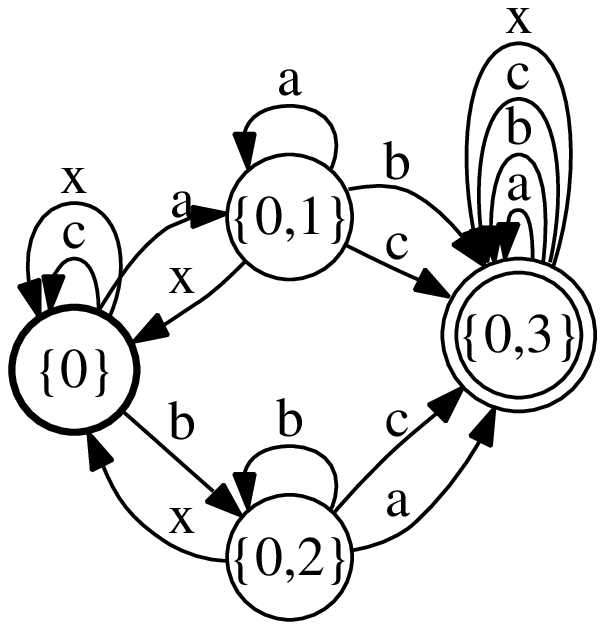}&
\ipsfig{.4}{figure=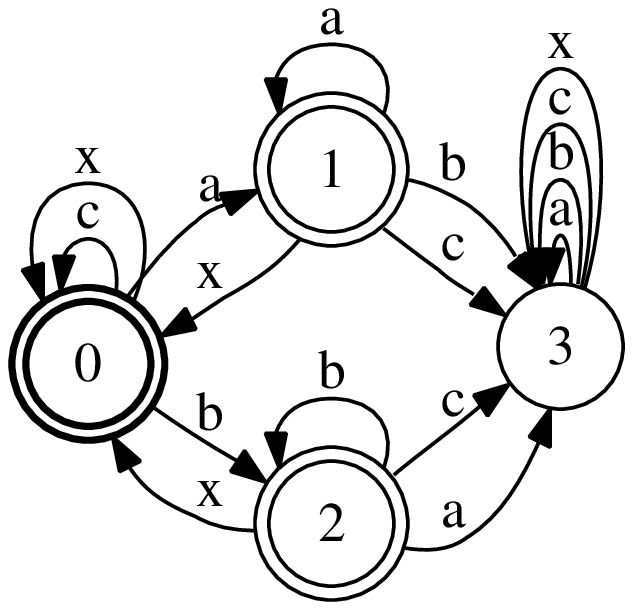}\\
(a) & (b) & (c)
\end{tabular*}
\end{center}
\caption{(a) Finite automaton $A$ representing the set of disallowed
sequences. (b) Automaton $B$, result of the determinization of
$A$. Subsets are indicated at each state. (c) Automaton $C$ obtained
from $B$ by complementation, state $3$ is not coaccessible.}
\label{fig:filter}
\end{figure}

\begin{proof}
Let $a$ denote $(\e_1 \coln \e_1)$, $b$ denote $(\e_2 \coln \e_2)$,
$c$ denote $(\e_2 \coln \e_1)$, and let $x$ stand for any $(x \coln
x)$, with $x \in \Sigma$. The following sequences must be disallowed
by a shortest-path filter with matching transitions first: $ab, ba,
ac, bc$.  This is because, from any state, instead of the moves $ab$
or $ba$, the matching or diagonal transition $c$ can be
taken. Similarly, instead of $ac$ or $bc$, $ca$ and $cb$ can be taken
for an earlier match. Conversely, it is clear from the grid or an
immediate recursion that a filter disallowing these sequences accepts
a unique path between two connected states of the grid.

Let $L$ be the set of sequences over $\sigma = \set{a, b, c, x}$ that
contain one of the disallowed sequence just mentioned as a substring
that is $L = \sigma^* (ab + ba + ac + bc)\sigma^*$. Then
$\overline{L}$ represents exactly the set of paths allowed by that
filter and is thus a regular language. Let $A$ be an automaton
representing $L$ (Figure~\ref{fig:filter}(a)). An automaton
representing $\overline{L}$ can be constructed from $A$ by
determinization and complementation (Figures~\ref{fig:filter}(a)-(c)).
The resulting automaton $C$ is equivalent to the transducer $M$
after removal of the state $3$, which does not admit a path to a final
state.\qed
\end{proof}
Thus, to compose two transducers $T_1$ and $T_2$ with
$\e$-transitions, it suffices to compute $\tilde{T}_1 \circ M \circ
\tilde{T}_2$, using the rules of composition in the $\e$-free case.

The problem of avoiding the creation of redundant $\e$-paths is more
complex in 3-way composition since the $\e$-transitions of all three
transducers must be taken into account. We describe two solutions for
this problem, one based on two filters, another based on a single
filter.

\subsection{$2$-way $\e$-Filters.}

\begin{figure}[t]
\begin{center}
\begin{tabular*}{.9\textwidth}{@{\extracolsep{\fill}}c@{\extracolsep{\fill}}c@{\extracolsep{\fill}}c@{\extracolsep{\fill}}c@{\extracolsep{\fill}}c@{\extracolsep{\fill}}}
\ipsfig{.4}{figure=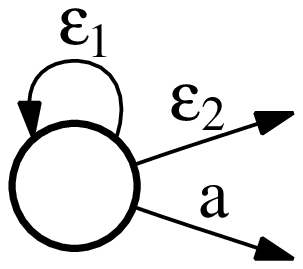}&
\ipsfig{.4}{figure=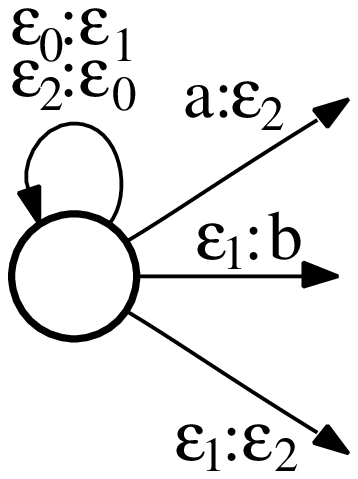}&
\ipsfig{.4}{figure=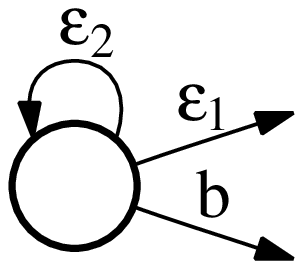}&
\ipsfig{.4}{figure=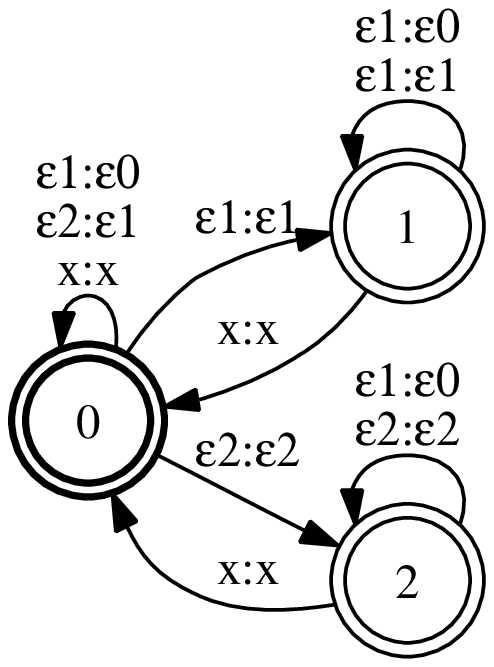}&
\ipsfig{.4}{figure=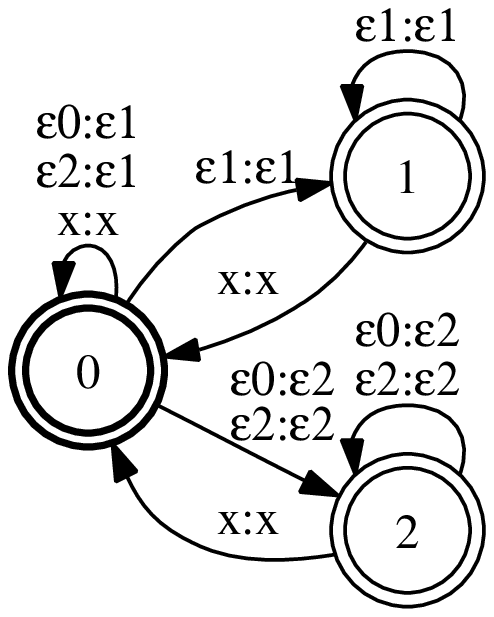}\\
(a) & (b) & (c) & (d) & (e)
\end{tabular*}
\end{center}
\caption{Marking of transducers and 2-way filters. (a)
$\tilde{T}_1$. Self-loop labeled with $\e_1$ added at all states of
$T_1$, regular output $\e$s renamed to $\e_2$. (b)
$\tilde{T}_2$. Self-loops with labels $(\e_0 \coln\e_1)$ and $(\e_2
\coln \e_0)$ added at all states of $T_2$. Input $\e$s are replaced by
$\e_1$, output $\e$s by $\e_2$. (c) $\tilde{T}_3$. Self-loop labeled
with $\e_2$ added at all states of $T_3$, regular input $\e$s renamed
to $\e_1$. (d) Left-to-right filter $M_1$. (e) Left-to-right filter
$M_2$. }
\label{fig:marking}
\end{figure}

One way to deal with this problem is to use the 2-way filter $M$, by
first dealing with matching $\e$-paths in $U = (T_1 \circ T_2)$, and
then $U \circ T_3$. However, in 3-way composition, it is possible to
remain at the same state of $T_1$ and the same state of $T_2$, and
move on an $\e$-transition of $T_3$, which previously was not an
option. This corresponds to staying at the same state of $U$, while
moving on a transition of $T_3$ with input $\e$. To account for this
move, we introduce a new symbol $\e_0$ matching $\e_1$ in $T_3$. But,
we must also ensure the existence of a self-loop with output label
$\e_0$ at all states of $U$. To do so, we augment the filter $M$ with
self-loops $(\e_1 \coln \e_0)$ and the transducer $T_2$ with
self-loops $(\e_0 \coln \e_1)$ (see
Figure~\ref{fig:marking}(b)). Figure~\ref{fig:marking}(d) shows the
resulting filter transducer $M_1$. From
Figures~\ref{fig:marking}(a)-(c), it is clear that $\tilde{T}_1 \circ
M_1 \circ \tilde{T}_2$ will have precisely a self-loop labeled with
$(\e_1 \coln \e_1)$ at all states.

In the same way, we must allow for moving forward on a transition of
$T_1$ with output $\e$, that is consuming $\e_2$, while remaining at
the same states of $T_2$ and $T_3$. To do so, we introduce again a new
symbol $\e_0$ this time only relevant for matching $T_2$ with $T_3$,
add self-loops $(\e_2 \coln \e_0)$ to $T_2$, and augment the filter
$M$ by adding a transition labeled with $(\e_0 \coln \e_2)$ (resp.
$(\e_0 \coln \e_1)$) wherever there used to be one labeled with $(\e_2
\coln \e_2)$ (resp. $(\e_2 \coln \e_1)$). Figure~\ref{fig:marking}(e)
shows the resulting filter transducer $M_2$.

Thus, the composition $\tilde{T}_1 \circ M_1 \circ \tilde{T}_2 \circ
M_2 \circ \tilde{T}_3$ ensures the uniqueness of matching $\e$-paths.
In practice, the modifications of the transducers $T_1$, $T_2$, and
$T_3$ to generate $\tilde{T}_1$, $\tilde{T}_2$, and $\tilde{T}_3$, as
well as the filters $M_1$ and $M_2$ can be directly simulated or
encoded in the 3-way composition algorithm for greater efficiency.
The states in $T$ become quintuples $(q_1, q_2, q_3, f_1, f_2)$ with
$f_1$ and $f_2$ are states of the filters $M_1$ and $M_2$.  The
introduction of self-loops and marking of $\e$s can be simulated (line
12-13) and the filter states $f_1$ and $f_2$ taken into account to
compute the set $G$ of the transition matches allowed (line 13).

Note that while 3-way composition is symmetric, the analysis of
$\e$-paths just presented is left-to-right and the filters $M_1$ and
$M_2$ are not symmetric. In fact, we could similarly define
right-to-left filters $M'_1$ and $M'_2$. The advantage of the filters
presented in this section is however that they can help modify easily
an existing implementation of composition into 3-way composition. The
filters needed for the 3-way case are also straightforward
generalizations of the $\e$-filter used in standard composition.

\subsection{$3$-way $\e$-Filter.}

There exists however a direct and symmetric method for dealing with
$\e$-paths in 3-way composition. Remarkably, this can be done using a
single filter automaton whose labels are 3-dimensional vectors.
Figure~\ref{fig:3-way} shows a filter $W$ that can be used for that
purpose. Each transition is labeled with a triplet. The $i$th element
of the triplet corresponding to the move on the $i$th transducer. $0$
indicates staying at the same state or not moving, $1$ that a move is
made reading an $\e$-transition, and $x$ a move along a matching
transition with a non-empty symbol (i.e., non-$\e$ output in $T_1$,
non-$\e$ input or output in $T_2$ and non-$\e$ input in $T_3$).

Matching $\e$-paths now correspond to a three-dimensional grid, which
leads to a more complex word combinatorics problem. As in the
two-dimensional case, $(p, q, r) \preceq (s, t, u)$ indicates that
$(s, t, u)$ can be reached from $(p, q, r)$ in the grid. Several
filters are possible, here we will again favor the matching of
$\e$-transitions (i.e. the diagonals on the grid).

\begin{proposition}
\label{prop:prop2}
The filter automaton $W$ allows a unique path between any two states
$(p, q, r)$ and $(s, t, u)$ of a three-dimensional grid, with $(p, q,
r) \preceq (s, t, u)$.
\end{proposition}

\begin{proof}
Let $\mathfrak M$ and $\mathfrak X$ be the defined by $\mathfrak M =
\set{(m_1, m_2, m_3): m_1, m_2, m_3 \in \set{0, 1}}$ and $\mathfrak X
= \set{(x, x, m), (m, x, x): m \in \set{0,1}}$. A sequence of moves
corresponding to a matching $\e$-path is thus an element of
$(\mathfrak M \cup \mathfrak X)^*$. Two sequences $\pi_1$ and $\pi_2$
are equivalent if they consume the same sequence of transitions on 
each of the three
transducers, for example $(0, x, x) (1, 1, 0)$ is equivalent to $(1,
x, x) (0, 1, 0)$. For each set of equivalent move sequences between
two states $(p, q, r)$ and $(s,t,u)$, we must preserve a unique
sequence representative of that set. We now define the unique
corresponding representative $\bar \pi$ of each sequence $\pi \in
(\mathfrak M \cup \mathfrak X)^*$. In all cases, $\overline{\pi}$ will
be the sequence where the $1$-moves and the $x$-moves are taken as early
as possible.

\begin{figure}[t]
\begin{center}
\ipsfig{.35}{figure=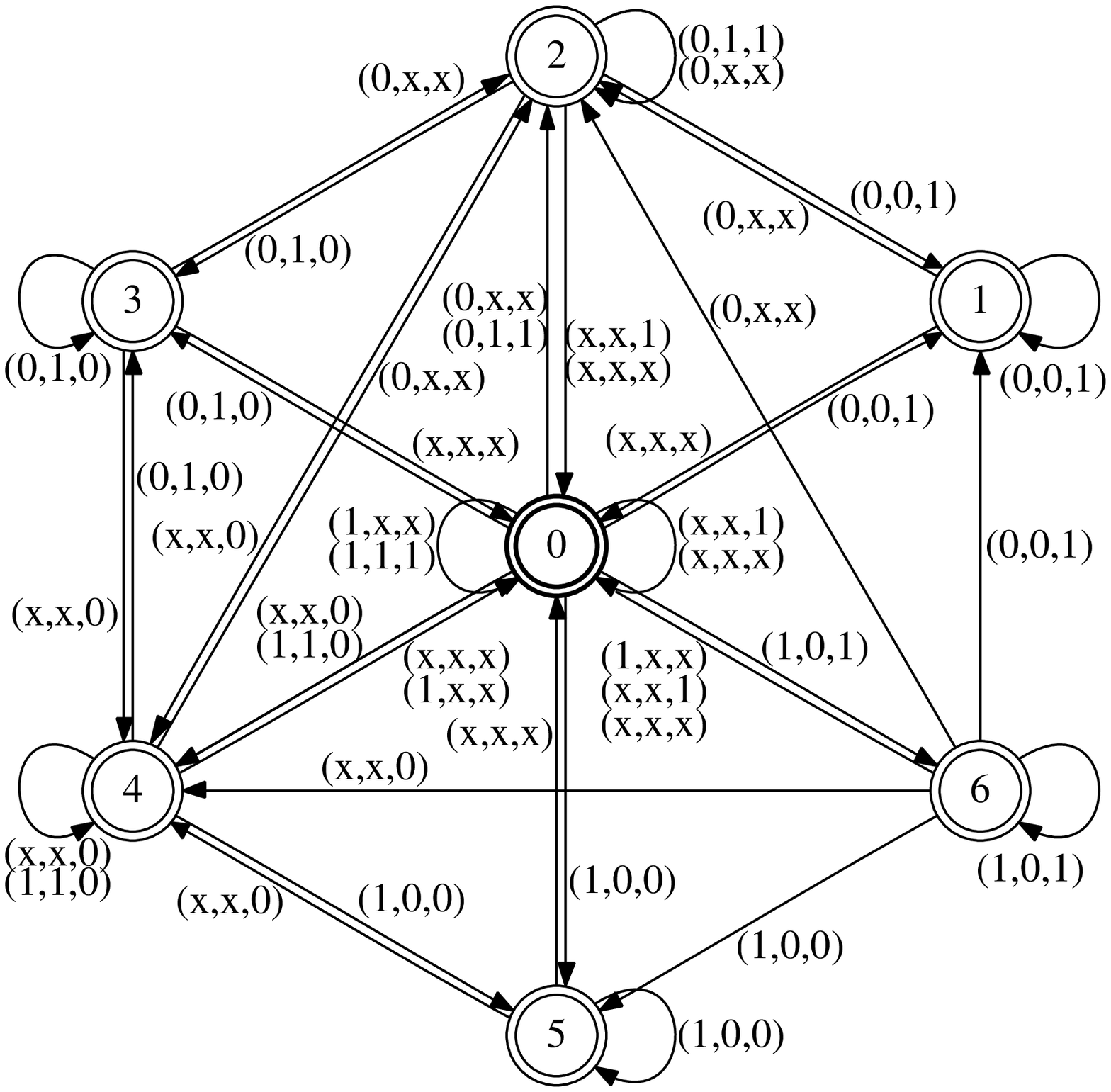}
\end{center}
\caption{3-way matching $\e$-filter $W$.}
\label{fig:3-way}
\end{figure}

\begin{enumerate}
\item Assume that $\pi \in \mathfrak M^*$ and let $n_i$ be the number
of occurrences of 1 as the $i$th element in a triplets defining
$\pi$. By symmetry, we can assume without loss of generality that $n_1
\le n_2 \le n_3$. We define $\overline{\pi}$ as $(1,1,1)^{n_1}
(0,1,1)^{n_2-n_1 }(0,0,1)^{n_3-n_2}$, that is the sequence where the
$1$-moves are taken as early as possible.

\item Otherwise, $\pi$ can be decomposed as $\pi = \mu_1 \chi_1 \mu_2
\chi_2 \cdots \mu_k \chi_k \mu_{k+1}$ with $k \ge 1$, $\mu_i \in
\mathfrak M^*$ and $\chi_i \in \mathfrak X$. $\overline{\pi}$ is then
defined by induction on $k$. By symmetry, we can assume that $\chi_1 =
(x,x,m)$ with $m \in \set{0,1}$. Let $\pi'$ be such that $\pi = \mu_1
\chi_1 \pi'$, let $n_i$ be the number of times 1 appears as $i$th
element in a triplet of $\mu_1$, and let $n_3'$ the number of times 1
is found as third element in a triplet reading $\chi_1 \pi'$ from left
to right before seeing an $x$.

\begin{enumerate}
\item If $n_3 \le \max(n_1, n_2)$, let $n = \min(n_3', \max(n_1, n_2)
- n_3)$. We can then obtain $\chi_1' \pi''$ by replacing the $n$ first
1's that appears in $\chi_1 \pi'$ as third element of a triplet by
0's. Let $\mu_1' = \mu_1 (0,0,1)^n$.  We then have that $\pi$ is
equivalent to $\mu_1' \chi_1' \pi''$. By induction, we can compute
$\overline{\pi''}$ and $\overline{\mu_1'}$ and define $\overline{\pi}$
as $\overline{\mu_1'} \chi_1' \overline{\pi''}$.
\item If $n_3 > \max(n_1, n_2)$, we define $n$ as $n_3 - \max(n_1,
n_2)$ if $\chi_1 = (x,x,1)$ and $n_3 - \max(n_1, n_2) -1 $ if $\chi_1
= (x,x,0)$. Let $\mu_1'$ be $(1,1,1)^{n_1} (0,1,1)^{n_2 - n_1}$ if
$n_1 < n_2$ and $(1,1,1)^{n_2} (1,0,1)^{n_1 - n_2}$ otherwise. We can
then define $\overline{\pi}$ as $\mu_1' (x,x,1) \overline{(0,0,1)^n
\pi'}$.
\end{enumerate}
\end{enumerate}
A key property of $\overline{\pi}$ is that it can be characterized by
a small set of forbidden sequences. Indeed, observe that the following
rules apply:
\begin{enumerate}
\item in two consecutive triplets, for $i \in [1, 3]$, 0 in the $i$th
machine of the first triplet cannot be followed by 1 in the second.
Indeed, as in the 2-way case, if we stay at a state, then we must
remain at that state until a match with a non-empty symbol is made
(this correspond to cases 1 and 2(a) of the definition of
$\overline{\pi}$).
\item two 0s in adjacent transducers ($T_1$ and $T_2$, or $T_2$ and
$T_3$), cannot become both $x$s unless all components become $x$s; For
example, the sequence $(0, 0, 1)(x, x, 1)$ is disallowed since instead
$(x, x, 1)(0, 0, 1)$ with an earlier match can be followed. Similarly,
the sequence $(0, 0, 1)(x, x, 0)$ is disallowed since instead the
single and shorter move $(x, x, 1)$ can be taken (this correspond to
case 2(b) of the definition).
\item the triplet (0, 0, 0) is always forbidden since it corresponds
to remaining at the same state in all three transducers.
\end{enumerate}
Conversely, we observe that with our definition of $\bar \pi$, these
conditions are also sufficient. Thus, a filter can be obtained by
taking the complement of an automaton accepting exactly the sequences
of forbidden substrings just described. The resulting deterministic
and minimal automaton is the filter $W$ shown in
Figure~\ref{fig:3-way}. \ignore{Conversely, it is not hard to see
that a filter disallowing these sequences accepts a unique path
between two connected states of the grid. It is not hard to see that
$\overline{\pi}$ will be the unique path not being disallowed by these
rules.} Observe that each state of $W$ has a transition labeled
by $(x,x,x)$ going to the initial state $0$, this corresponds to
resetting the filter at the end of a matching $\e$-path. \qed 
\end{proof}

The filter $W$ is used as follows. A triplet state $(q_1, q_2, q_3)$
in $3$-way composition is augmented with a state $r$ of the filter
automaton $W$, starting with state $0$ of $W$. The transitions of the
filter $W$ at each state $r$ determine the matches or moves allowed
for that state $(q_1, q_2, q_3, r)$ of the composed machine.

\section{Experiments}
\label{sec:experiments} 

This section reports the results of experiments carried out in two
different applications: the computation of a complex edit-distance
between two automata, as motivated by applications in text and speech
processing \cite{edit}, and the computation of kernels between
automata needed in spoken-dialog classification and other machine
learning tasks.

\begin{table}[h]
\label{table:exp}
\caption{Comparison of 3-way composition with standard
composition. The computation times are reported in seconds, the size
of $T_2$ in number of transitions. These experiments were performed on
a dual-core AMD Opteron 2.2GHz with 16GB of memory, using the same
software library and basic infrastructure.}
\begin{center}
\begin{tabular}{l@{\hspace{.5cm}}rrrrrr@{\hspace{.5cm}}r@{\hspace{.3cm}}r}
\hline\noalign{\smallskip} 
 & \multicolumn{6}{c}{$n$-gram Kernel} & \multicolumn{2}{c}{Edit distance}\\
 & $\le 2$ & $\le 3$ & $\le 4$ & $\le 5$ & $\le 6$ & $\le 7$ & standard & +transpositions \\
\noalign{\smallskip} 
\hline 
\noalign{\smallskip} 
Standard & 65.3 & 68.3 & 71.0 & 73.5 & 76.3 & 78.3 & 586.1 & 913.5\\
3-way & 8.0 & 8.1 & 8.2 & 8.2 & 8.2 & 8.2 & 3.8 & 5.9\\
\hline
Size of $T_2$ & 70K & 100K & 130K & 160K & 190K & 220K & 25M & 75M \\ 
\end{tabular}
\end{center}
\vspace{-.75cm}
\end{table}

In the edit-distance case, the standard transducer $T_2$ used was one
based on all insertions, deletions, and substitutions with different
costs \cite{edit}. A more realistic transducer $T_2$ was one augmented
with all transpositions, e.g., $ab \rightarrow ba$, with different
costs. In the kernel case, $n$-gram kernels with varying $n$-gram
order were used \cite{jmlr}.

Table~\ref{table:exp} shows the results of these experiments. The
finite automata $T_1$ and $T_3$ used were extracted from real text and
speech processing tasks. The results show that in all cases, 3-way
composition is orders of magnitude faster than standard composition.

\section{Conclusion}
\label{sec:conclusion}

We presented a general algorithm for the composition of weighted
finite-state transducers. In many instances, 3-way composition
benefits from a significantly better time and space complexity. Our
experiments with both complex edit-distance computations arising in a
number of applications in text and speech processing, and with kernel
computations, crucial to many machine learning algorithms applied to
sequence prediction, show that our algorithm is also substantially
faster than standard composition in practice. We expect 3-way
composition to further improve efficiency in a variety of other areas
and applications in which weighted composition of transducers is used.

\subsubsection*{Acknowledgments.}
{\footnotesize
The research of Cyril Allauzen and Mehryar Mohri was partially
supported by the New York State Office of Science Technology and
Academic Research (NYSTAR). This project was also sponsored in part by
the Department of the Army Award Number W81XWH-04-1-0307. The
U.S. Army Medical Research Acquisition Activity, 820 Chandler Street,
Fort Detrick MD 21702-5014 is the awarding and administering
acquisition office.  The content of this material does not necessarily
reflect the position or the policy of the Government and no official
endorsement should be inferred.
}
\bibliographystyle{plain}
\bibliography{3way}

\begin{thebibliography}{10}

\bibitem{berstel}
Jean Berstel.
\newblock {\em Transductions and Context-Free Languages}.
\newblock Teubner, 1979.

\bibitem{Chen98}
Stanley Chen and Joshua Goodman.
\newblock An empirical study of smoothing techniques for language modeling.
\newblock Technical Report, TR-10-98, Harvard University, 1998.

\bibitem{jmlr}
Corinna Cortes, Patrick Haffner, and Mehryar Mohri.
\newblock {Rational Kernels: Theory and Algorithms}.
\newblock {\em Journal of Machine Learning Research}, 5:1035--1062, 2004.

\bibitem{Culik1997}
Karel {Culik II} and Jarkko Kari.
\newblock {Digital Images and Formal Languages}.
\newblock In Grzegorz Rozenberg and Arto Salomaa, editors, {\em Handbook of
  Formal Languages}, volume~3, pages 599--616. Springer, 1997.

\bibitem{eilenberg}
Samuel Eilenberg.
\newblock {\em Automata, Languages and Machines}.
\newblock Academic Press, 1974--76.

\bibitem{Katz87}
Slava~M. Katz.
\newblock Estimation of probabilities from sparse data for the language model
  component of a speech recogniser.
\newblock {\em IEEE Transactions on Acoustic, Speech, and Signal Processing},
  35(3):400--401, 1987.

\bibitem{kuich}
Werner Kuich and Arto Salomaa.
\newblock {\em Semirings, Automata, Languages}.
\newblock Number~5 in {EATCS} Monographs on Theoretical Computer Science.
  Springer-Verlag, 1986.

\bibitem{cl}
Mehryar Mohri.
\newblock {Finite-State Transducers in Language and Speech Processing}.
\newblock {\em Computational Linguistics}, 23(2), 1997.

\bibitem{edit}
Mehryar Mohri.
\newblock {Edit-Distance of Weighted Automata: General Definitions and
  Algorithms}.
\newblock {\em International Journal of Foundations of Computer Science},
  14(6):957--982, 2003.

\bibitem{lothaire}
Mehryar Mohri.
\newblock {Statistical Natural Language Processing}.
\newblock In M.~Lothaire, editor, {\em {Applied Combinatorics on Words}}.
  Cambridge University Press, 2005.

\bibitem{ecai}
Mehryar Mohri, Fernando C.~N. Pereira, and Michael Riley.
\newblock {Weighted Automata in Text and Speech Processing}.
\newblock In {\em Proceedings of the 12th biennial European Conference on
  Artificial Intelligence (ECAI-96)}. John Wiley and Sons, 1996.

\bibitem{pereira-riley}
Fernando Pereira and Michael Riley.
\newblock {\em Finite State Language Processing}, chapter Speech Recognition by
  Composition of Weighted Finite Automata.
\newblock The MIT Press, 1997.

\bibitem{perrin-lothaire}
Dominique Perrin.
\newblock Words.
\newblock In M.~Lothaire, editor, {\em Combinatorics on words}, Cambridge
  Mathematical Library. Cambridge University Press, 1997.

\bibitem{soittola}
Arto Salomaa and Matti Soittola.
\newblock {\em {A}utomata-{T}heoretic {A}spects of {F}ormal {P}ower {S}eries}.
\newblock Springer-Verlag, 1978.

\end{thebibliography}
\end{document}